\documentclass{mem}
\usepackage{natbib}\usepackage{txfonts}\usepackage{balance}
\usepackage{graphicx}
\usepackage[a4paper,breaklinks,dvipdfm]{hyperref}
\idline{75}{282}
\begin{document}
\def\teff{$T\rm_{eff }$}
\def\kms{$\mathrm {km s}^{-1}$}

\title{
Kinematics of Star Clusters in M101
}

   \subtitle{}

\author{
L. \,Simanton\inst{1} 
\and R. \, Chandar\inst{1}
\and B. \, Miller\inst{2}
          }

  \offprints{L. Simanton}

\institute{
University of Toledo, Department of Physics and Astronomy
2801 W. Bancroft St,
Toledo, OH 43606
\email{lesley.simanton@utoledo.edu}
\email{rupali.chandar@utoledo.edu}
\and
Gemini Observatory
25 Long Street, 255255,
La Serena, Chile 
\email{bmiller@gemini.edu}
}

\authorrunning{Simanton }

\titlerunning{M101 Clusters}

\abstract{
We have identified a few thousand star clusters in the nearby, late-type spiral galaxy M101, including $\approx$~90 candidate ancient globular clusters (GCs), from multi-band \textit{Hubble Space Telescope} ($HST$) images.  We obtained follow-up low-resolution ($R \approx$~2000) optical spectroscopy from Gemini-GMOS for 43 total clusters, of which 18 are old GCs and 25 are young massive clusters (YMCs).  We measure radial velocities for these clusters and find that, as expected, the YMCs rotate with the HI disk.  The old GCs do not show any obvious evidence for rotation and have a much higher velocity dispersion than the YMCs, suggesting that the GCs in M101 are likely part of a stellar halo or thick disk.
}
\maketitle{}

\section{Introduction}

The old GCs in the Milky Way are associated with the halo and the thick disk \citep{car07}, but not the thin disk.  However, GCs in some late-type galaxies, such as NGC 253 \citep{ols04} and the LMC \citep{fre83}, appear to rotate with the thin disk.  Young clusters meanwhile are associated with the thin disk in spirals and late-type galaxies.  Here, we look at the kinematics of GCs and YMCs in the late-type spiral M101 to determine the structure of its cluster population.

\section{$HST$ Cluster Candidates and Follow-up GMOS Spectra} 

Using $HST$ BVI images, we identified $\approx$~90 candidate GCs and $\approx$~4,000 YMC candidates.  Cluster candidates were selected to be broader than the point spread function.  We selected a subsample of clusters brighter than V$\leq$~21.5 for follow-up spectroscopy (see Fig. 1 and 2).

Follow-up spectroscopy was obtained using the $Gemini$ Multi-Object Spectrograph (GMOS) North instrument with the B600-5303 grating.  Spectra were categorized visually as either GCs or YMCs.  See Figures 3 and 4 for typical spectra.  In total, 43 clusters were observed, 18 of which were identified as GCs and 25 as YMCs.  

\begin{figure*}[t!]
\resizebox{12.5cm}{!}{\includegraphics[clip=true]{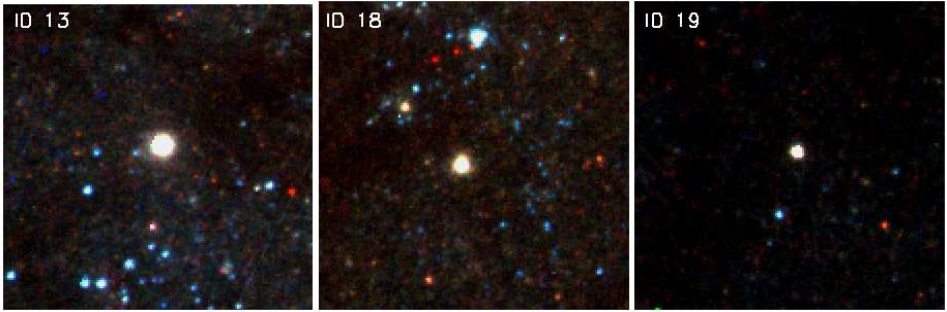}}
\caption{\footnotesize Close-ups of GCs from $HST$ ACS color-combined $BVI$ images.  Each postage stamp is approximately 7.35" $\times$ 7.35".}
\label{YMC}
\end{figure*}

\begin{figure*}[t!]
\resizebox{12.5cm}{!}{\includegraphics[clip=true]{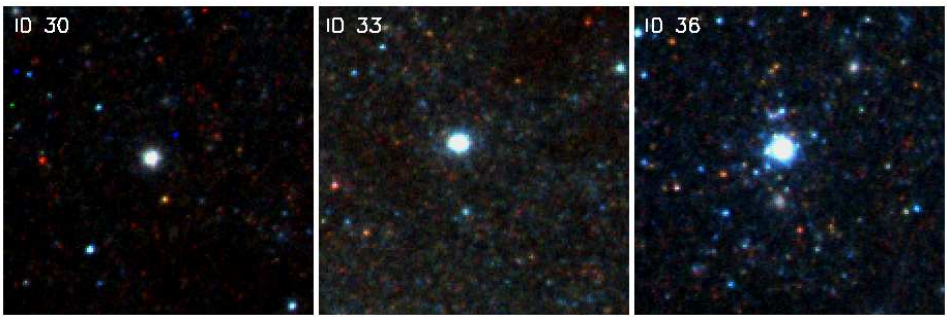}}
\caption{\footnotesize YMCs close-ups with the same specifications as Figure 1. }
\label{YMC}
\end{figure*}

\begin{figure*}[t!]
\resizebox{\hsize}{!}{\includegraphics[clip=true]{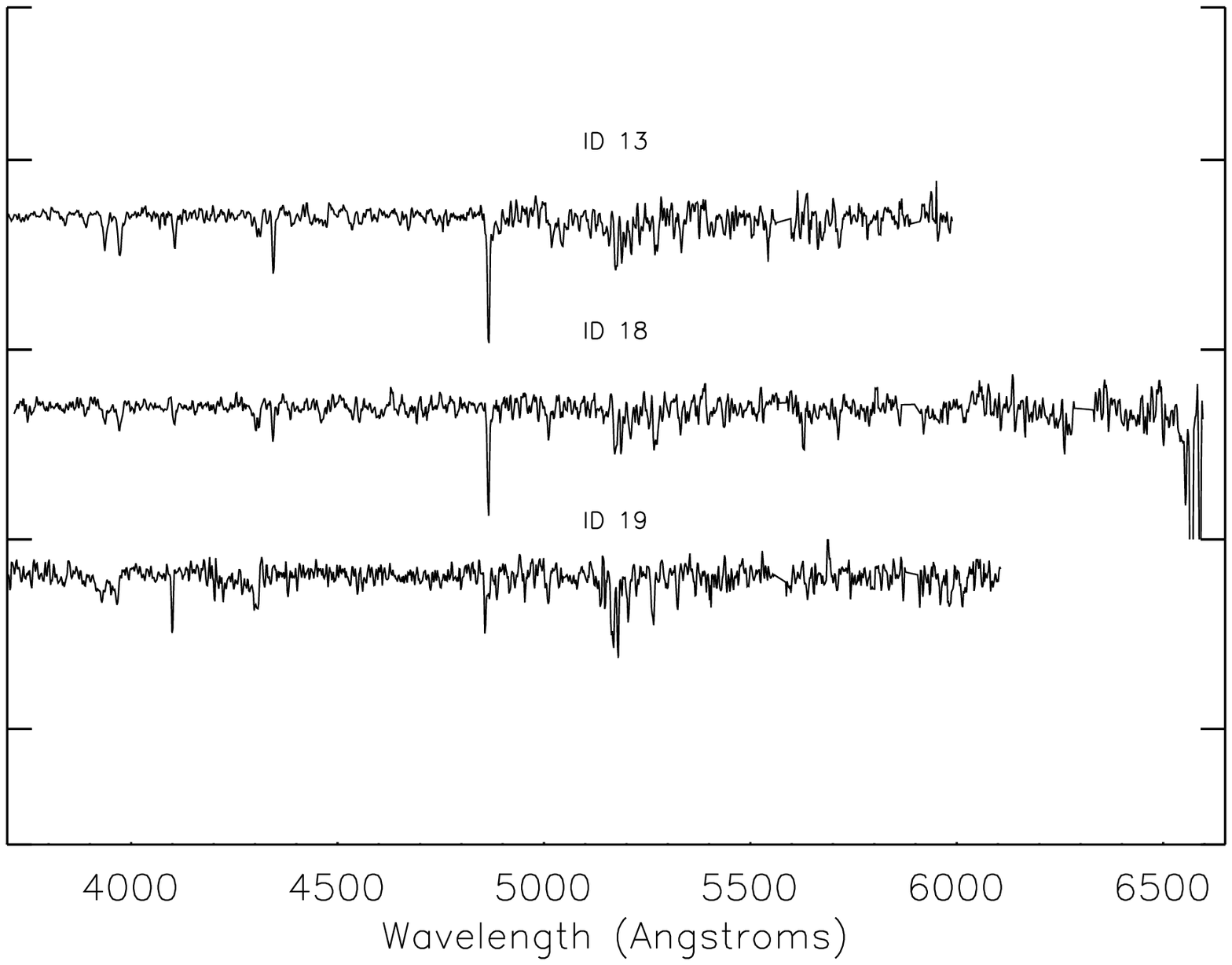}}
\caption{\footnotesize GMOS spectra of GCs reduced with methods similar to those in appendix A2 of \citet{tra07} and continuum normalized.  GCs were identified by their weak Balmer lines and noticable Ca H \& K, G band, and Fe lines.}
\resizebox{\hsize}{!}{\includegraphics[clip=true]{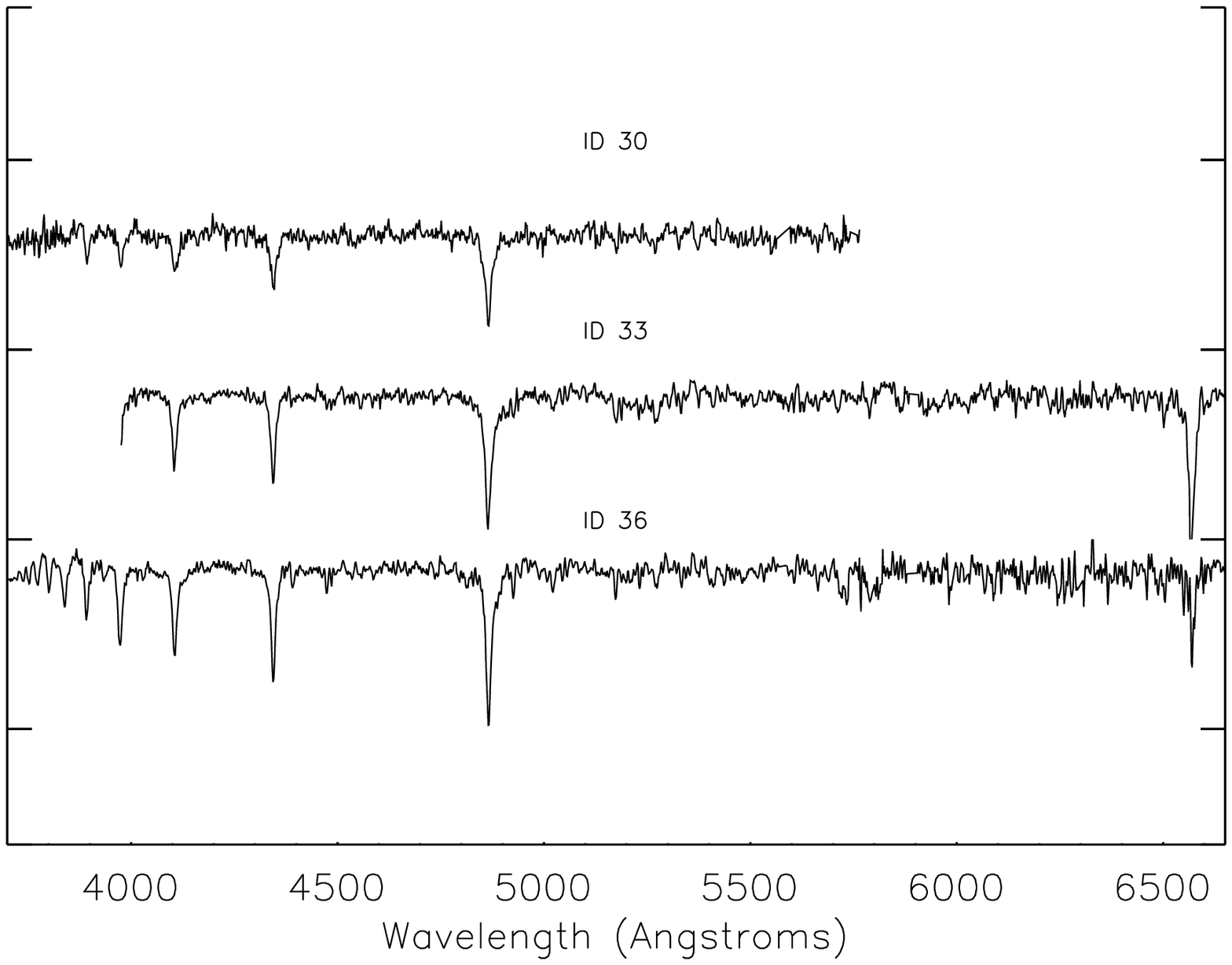}}
\caption{\footnotesize GMOS spectra of YMCs with the same specifications as Figure 3.  YMCs were identified by their strong Balmer lines and lack of the G band.}
\label{YMC}
\end{figure*}

\section{Kinematic Results}

We measured the velocities of our clusters by cross-correlation with template spectra and found the distance from each cluster to the semi-minor axis (see Fig. 5).  We then plotted the velocities versus the distances in Figure 6 with the best fit line for the GCs and YMCs shown separately.  Also plotted is the best fit line for the disk velocities, which were found from the HI maps in \citet{bos81}.  

\begin{figure*}[t!]
\resizebox{\hsize}{!}{\includegraphics[clip=true]{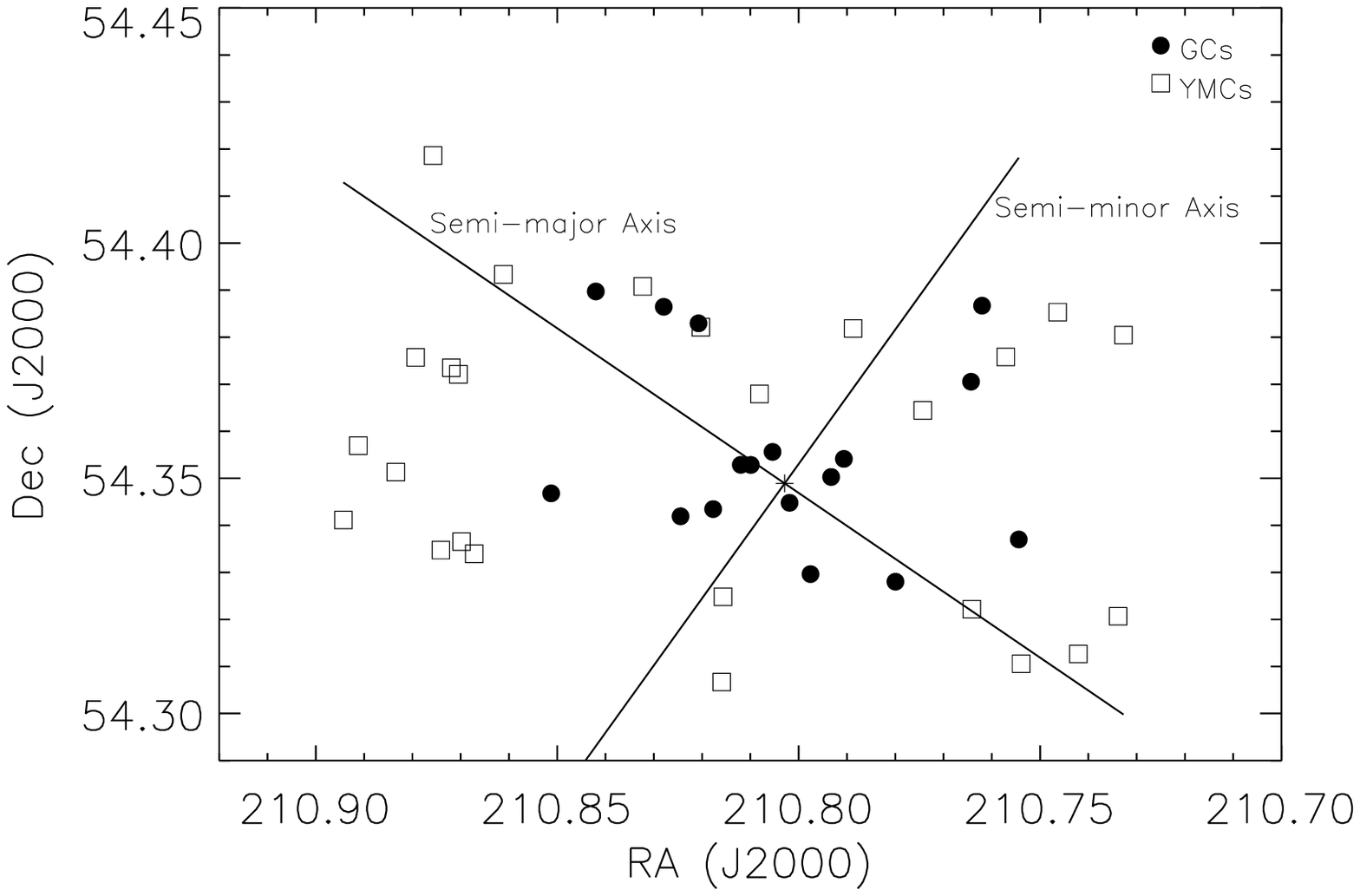}}
\caption{\footnotesize Positions for GCs (filled circles) and YMCs (open squares) as compared to the semi-minor axis.}
\resizebox{\hsize}{!}{\includegraphics[clip=true]{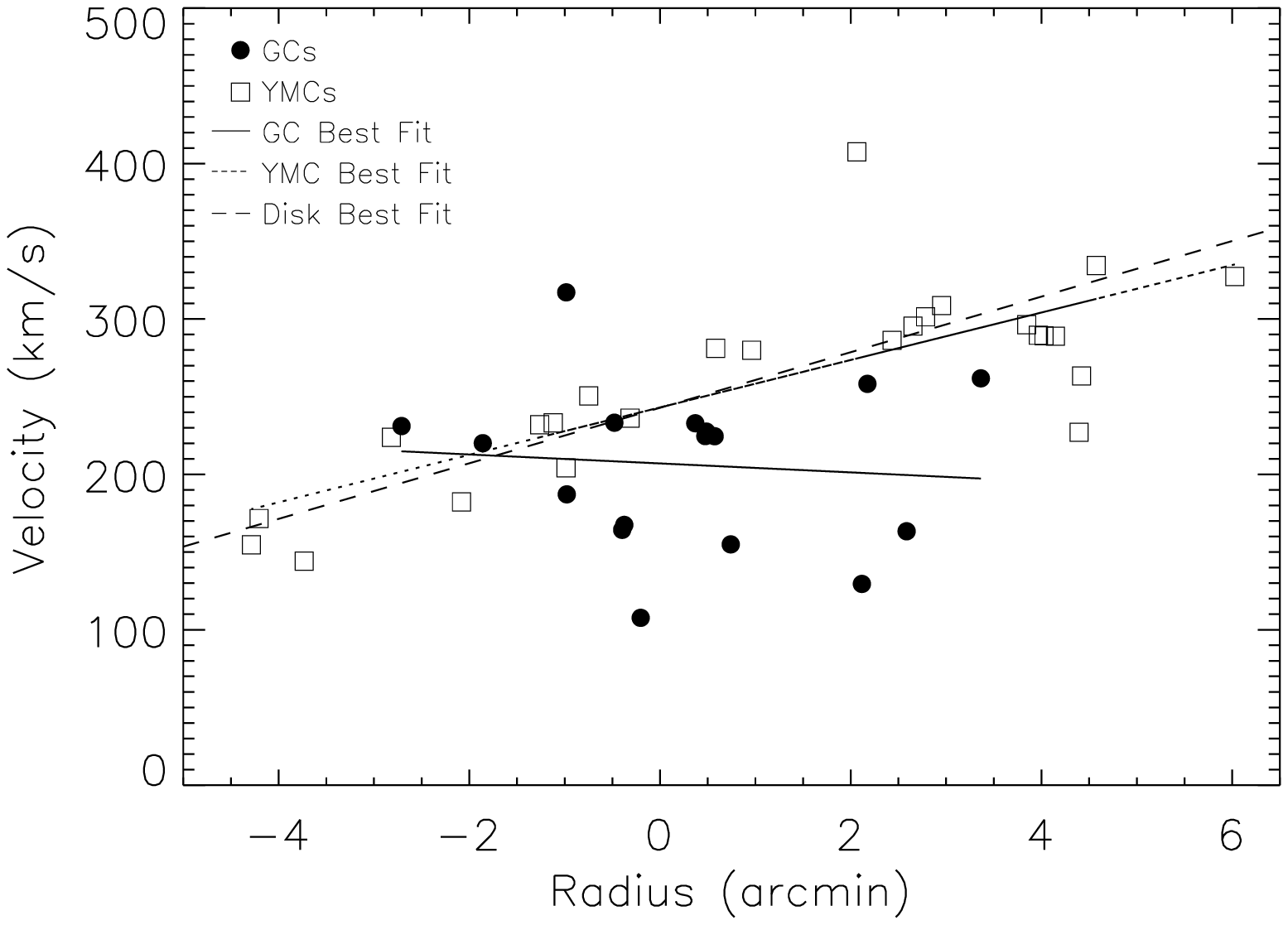}}
\caption{\footnotesize Velocities versus radius for GCs (filled circles) and YMCs (open squares) where the ``radius" is the perpendicular distance from each cluster to the semi-minor axis (see Fig. 5).  The best fit lines for both cluster types are over plotted.  The best fit line for the HI gas of the disk is also shown, and it is clear that the GCs are not associated with the thin disk. }
\label{YMC}
\end{figure*}

It is clear from the slopes of the best fit lines that the GCs do not lie in the thin disk, whereas the YMC velocities follow the HI disk much more closely.  The velocity dispersion of the GCs ($\sigma_{GC} \approx 52$~km/s) being larger than that of the YMCs ($\sigma_{YMC} \approx 36$~km/s) also supports this conclusion. 

\section{Future Work}

The next step in analyzing the cluster population of M101 will be to measure quantitative strengths of absorption lines to more precisely determine the ages and metallicities of GCs in our sample.  We can then combine precise ages and metallicities from spectra with photometry from the $HST$ multi-band images of the entire GC population of M101 for a more detailed analysis including assessing whether the color/metallicity of the system is bimodal.


\bibliographystyle{aa}

\begin{thebibliography}{}

\bibitem[{Bosma et al. (1981)}]{bos81} 
Bosma, A., Goss, W.\ M., \& Allen, R.\ J. 1981, A\&A, 93, 106

\bibitem[{Carollo et al. (2007)}]{car07}
Carollo, D. et al. 2007, Nature, 450, 1020 

\bibitem[{Freeman et al. (1983)}]{fre83}
Freeman, K.\ C., Illingworth, G., \& Oemler, A., Jr. 1983, ApJ 272, 488

\bibitem[{Olsen et al. (2004)}]{ols04}
Olsen, K.\ A.\ G., et al. 2004, AJ, 127, 2674 

\bibitem[{Trancho et al. (2007)}]{tra07}
Trancho, G. et al.\ 2007, ApJ, 664, 284


\end{thebibliography}

\end{document}